\documentclass[11pt]{article}

\usepackage{graphicx}
\usepackage{color}
\usepackage{array}
\usepackage{rotating}
\usepackage{amsmath}
\usepackage{amssymb}
\usepackage{appendix}
\usepackage{upgreek}

\usepackage{multibib}
\newcites{main}{Main References}
\newcites{methods}{Methods References}

\usepackage{graphicx}
\usepackage{fullpage}
\usepackage{amsmath}
\usepackage[version=3]{mhchem} 

\def\E{\mathrm{e}}
\usepackage{authblk}
\usepackage{lineno}
\begin{document}

\title{Enabling martian habitability with silica aerogel via the solid-state greenhouse effect}
\author[1,2]{R.~Wordsworth}
\author[3]{L.~Kerber}
\author[4]{C.~Cockell}

\affil[1]{School of Engineering and Applied Sciences, Harvard University, Cambridge, MA, USA}
\affil[2]{Department of Earth and Planetary Sciences, Harvard University, Cambridge, MA, USA}
\affil[3]{Jet Propulsion Laboratory, California Institute of Technology, Pasadena, CA, USA}
\affil[4]{UK Centre for Astrobiology, School of Physics and Astronomy, University of Edinburgh, Edinburgh, UK}

\maketitle

\textbf{
The low temperatures \citemain{Martinez2017,Forget1999} and high ultraviolet (UV) radiation levels \citemain{Cockell2000} at the surface of Mars today currently preclude the survival of life anywhere except perhaps in limited subsurface niches \citemain{Michalski2018}. Several ideas for making the martian surface more habitable have been put forward previously \citemain{Allaby1984,McKay1991b,zubrin1993technological,gerstell2001keeping}, but they all involve massive environmental modification that will be well beyond human capability for the foreseeable future \citemain{Jakosky2018}. Here we present a new approach to this problem. We show that widespread regions of the surface of Mars could be made habitable to photosynthetic life in the future via a solid-state analogue to Earth's atmospheric greenhouse effect. Specifically, we demonstrate via experiments and modelling that under martian environmental conditions, a 2 to \mbox{3-cm} thick layer of silica (\ce{SiO2}) aerogel will simultaneously transmit sufficient visible light for photosynthesis, block hazardous ultraviolet radiation, and raise temperatures underneath permanently to above the melting point of water, without the need for any internal heat source. Placing silica aerogel shields over sufficiently ice-rich regions of the martian surface could therefore allow photosynthetic life to survive there with minimal subsequent intervention. This regional approach to making Mars habitable is much more achievable than global atmospheric modification. In addition, it can be developed systematically starting from minimal resources, and can be further tested in extreme environments on Earth today.
}\newline

Around 50~K of surface warming is required on Mars to raise annual average low to mid-latitude temperatures to above the melting point of liquid water.  Mars' current atmosphere is too thin to significantly attenuate UV or to provide greenhouse warming of more than a few degrees K. However, observations of dark spots on Mars' polar \ce{CO2} ice caps suggest that they are transiently warmed by a greater amount via a planetary phenomenon known as the solid-state greenhouse effect \citemain{Hapke1996,Kieffer2006,Pilorget2011,Kaufmann2017}, which arises when sunlight becomes absorbed in the interior of translucent snow or ice layers \citemain{Niederdorfer1933,Brown1987}. The solid-state greenhouse effect is strongest in materials that are partially transparent to visible radiation but have low thermal conductivity and low infrared transmissivity (Fig.~\ref{fig:schematic_1}).  While carbon dioxide and \ce{H2O} ices are common on Mars, they are much too volatile to make robust solid-state greenhouse shields for life. Silica (\ce{SiO2}) has more favorable properties in that it is chemically stable and refractory at martian surface temperatures. Solid silica is transparent to visible radiation but opaque to UV {at wavelengths shorter than} 200-400~nm and infrared {at wavelengths longer than} $\sim 2$~$\upmu$m, depending on the abundance of impurities such as --OH groups. However, at 0.8--1.6~W/m/K \citemain{CRC2000}, the thermal conductivity of solid silica is too high to allow a strong warming effect. 

Silica aerogels, which consist of nanoscale networks of interconnecting \ce{SiO2} clusters, are over 97\% air by volume and have some of the lowest measured thermal conductivities of any known material (around 0.02~W/m/K at 1~bar pressure or 0.01~W/m/K at martian atmospheric pressures) \citemain{Dorcheh2008}. Because of these properties, silica aerogels have gained prominence in many fields of engineering, including in the design of passively heated buildings on Earth \citemain{Schultz2005} and even in the Mars Exploration Rovers, where thin aerogel layers were used to provide night-time thermal insulation \citemain{Novak2003}. Silica aerogels therefore hold excellent potential for creating strong solid-state greenhouse warming under martian conditions.

We have performed experiments to demonstrate the warming potential of silica aerogel solid-state greenhouse layers under Mars-like insolation levels. Our experimental setup consists of a layer of silica aerogel particles or tiles (see~Fig.~\ref{fig:image}) on a low reflectivity base surrounded by thermally insulating material (see Methods). The apparatus is exposed to visible radiation from a solar simulator. The broadband flux incident on the layer is measured with a pyranometer, and temperature is recorded by calibrated glass-bead thermistors. 

Figure~\ref{fig:results} shows the experimental results for both aerogel particle and tile layers vs. received visible flux in the 100-200~W/m$^2$ range. For comparison, Earth's global mean received flux is 342~W/m$^2$, while that of Mars is 147~W/m$^2$. As can be seen, temperature differences of over 45~K  are achieved for aerogel particle layers of thickness 3~cm receiving a flux of 150~W/m$^2$. Aerogel tiles, which have higher visible transmission, cause temperature differences that are around 10~K higher, reaching $>50$~K at just 2~cm thickness. Our experimental results show that under Mars-like insolation levels, warming to the melting point of liquid water or higher can be obtained under a 2-3~cm thick silica aerogel layer. The peak obtainable warming is likely even higher (see Methods), because heat is lost in our experimental setup via sidewall and base thermal losses and convection. We also measured the transmission of the aerogel particles and tiles in the ultraviolet and found strong attenuation of UV-AB, and near-total attenuation of the most hazardous UV-C radiation (Fig.~\ref{fig:transmission}). 

While raising surface temperatures and blocking UV radiation are the two most critical considerations for permitting life to survive on Mars, additional constraints due to atmospheric pressure, nutrient availability and dust deposition also need to be considered. {Brines can remain liquid below the freezing point of pure water, which could lower the temperature requirement below the 273~K we have assumed here \citemain{Fairen2009}, although for high enough salinities habitability would become restricted to halophilic organisms only. The higher \ce{CO2} partial pressure on Mars versus Earth is favorable for plant growth \citemain{McKay1991b}, but the low total atmospheric pressure means that at temperatures of 273~K or higher, the undersides of silica aerogel greenhouse shields would need to remain slightly pressurized relative to the atmosphere to avoid loss of water vapour either vertically or laterally. This would place light demands on their structural properties, which could {plausibly} be met by interspersing the silica aerogel with thin layers of solid transparent material or via organic polymer reinforcement \citemain{Meador2005,Randall2011}. Most nutrients appear to be readily available on the martian surface, with the abundances of some (such as Fe and S) higher than on Earth \citemain{Nixon2013}. The low partial pressure of \ce{N2} on Mars may pose a challenge for nitrogen fixation by unadapted terrestrial microorganisms. However, nitrate deposits, which have been observed \emph{in situ} on the surface of Mars, are a plausible alternative source of N \citemain{Stern2015}. 

The most favorable locations on Mars for creating local life-supporting regions are those that combine the key resources of light and surface \ce{H2O} while minimizing hazards such as excessive dust deposition. Within the $\pm 45^\circ$ latitude band where solar flux is high throughout the year, there are many mid-latitude locations where observations indicate the presence of near-surface ground ice \citemain{Mustard2001,Plaut2009,Dundas2018} and  climate model simulations \citemain{Kahre2006} indicate dust accumulation rates should be low.  Figure~\ref{fig:sim_SSG} shows the results of a coupled radiative-thermal calculation for the evolution of martian subsurface temperatures at one such location (Deuteronilus Mensae) in the presence of a solid-state greenhouse silica aerogel layer. Our model takes into account changes in martian insolation and aerogel radiative transfer, thermal conduction both in the aerogel and the regolith, and the latent heat associated with melting/freezing of regolith ice (see Methods). As can be seen, given an assumed 2.5-cm thick aerogel layer, sub-surface temperatures down to several metres depth are high enough to allow liquid water throughout the martian year after a few years at this location.

Our results show that via the solid-state greenhouse effect, regions on the surface of Mars could be modified in the future to allow life to survive there with much less infrastructure or maintenance than via other approaches. The creation of permanently warm regions would have many benefits for future {human activity on Mars}, as well as being of fundamental interest for astrobiological experiments and as a potential means to facilitate life detection efforts \citemain{Schulze2013}. 
The solid-state greenhouse warming concept also has applications for research in hostile environments on Earth today,  such as Antarctica and Chile's Atacama desert. 

In future work, it will be important to investigate the ease with which traditional silica aerogel manufacturing techniques \citemain{Dorcheh2008} can be adapted to conditions on Mars. However, given the ability of life on Earth to modify its environment,  it is also interesting to consider the extent to which organisms could eventually contribute to sustaining martian habitable conditions themselves. On Earth, multiple organisms already exist that utilize silica as a building material, including hexactinellid sponges and diatom phytoplankton \citemain{Aizenberg2005,Kroger2008}.  Diatoms in particular can grow up to several mm in length, produce frustules from $\sim$1-10~nm diameter amorphous silica particles (smaller than the mean pore diameter in silica aerogel networks) \citemain{Dorcheh2008,Parkinson1999}, and are already known to have high potential for bionanotechology applications in other areas \citemain{Kroger2008,Nassif2011}. It could therefore be interesting in the future to investigate whether high visible transmissivity, low thermal conductivity silica layers could be produced directly via a synthetic biology approach. If this is possible, in combination with the results described here it 
could eventually allow the development of a self-sustaining biosphere on Mars.

The fact that making Mars habitable to photosynthetic life is a potentially achievable near- to medium-term goal raises important ethical and philosophical questions. Most obviously, if Mars still possesses extant life today, its survival or detection might be hampered by the presence of Earth-based microorganisms \citemain{Rummel2014}. However, no mission has yet detected life on Mars, so if it does exist, it is likely to be confined to very specific regions in the subsurface. The approach studied here would not result in the survival of Earth-based life outside of solid-state greenhouse regions, so it should be unlikely to pose a greater risk to the search for martian life than the presence of humans on the surface. Nonetheless, the planetary protection concerns surrounding the transfer of Earth-based life to Mars are important, so the astrobiological risks associated with this approach to enabling martian habitability will need to be weighed carefully against the benefits to Mars science and human exploration in future.

\textbf{Corresponding author:} All correspondence and requests for materials should be addressed to R.~Wordsworth. 
 
\textbf{Author contributions:} All authors discussed the experimental and numerical results and contributed to a draft version of the manuscript. R.~Wordsworth proposed the solid-state greenhouse habitability idea, performed the experiments, and wrote much of the manuscript. L.~Kerber helped to write the sections on site selection on Mars and silica aerogel properties, and compiled data on the evidence for subsurface water sources. C.~Cockell provided input on the nutrient, ultraviolet, and pressure constraints on martian habitability and on the ethical considerations. 

\newpage

\begin{figure}[h!]
	\begin{center}
		{\includegraphics[width=3.5in]{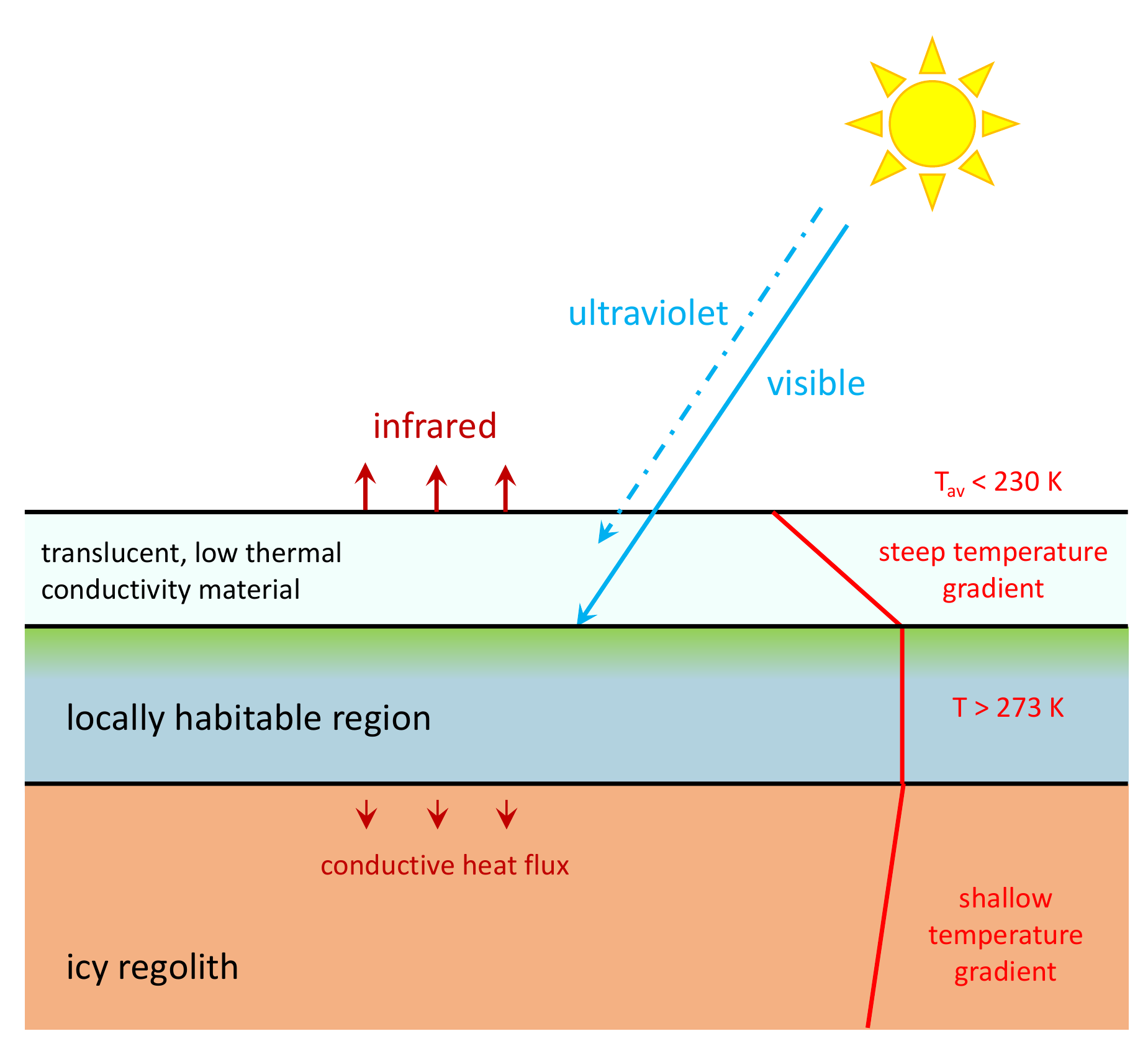}}
	\end{center}
	\caption{\textbf{Schematic of the solid-state greenhouse habitability concept for Mars.} A thin translucent layer of low thermal conductivity material transmits visible light but blocks ultraviolet and infrared, directly replicating the radiative effects of Earth's atmosphere. The depth of the habitable region in the subsurface increases with time due to thermal diffusion.}
\label{fig:schematic_1}
\end{figure}

\begin{figure}[h!]
	\begin{center}
		{\includegraphics[width=3.5in]{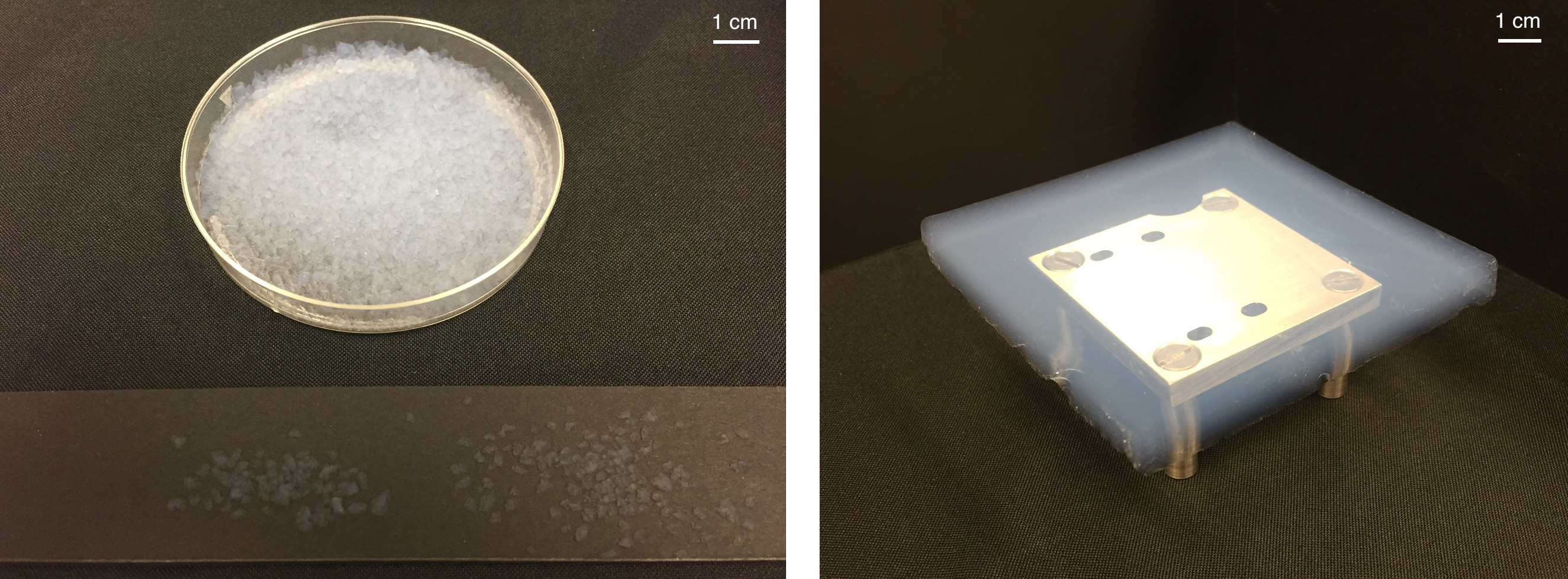}}
	\end{center}
	\caption{\textbf{Image of silica aerogel used in the experiment.} Left and right panels show the aerogel particles and an aerogel tile, respectively. In each case the white bar in the top right indicates the scale.}
\label{fig:image}
\end{figure}

\begin{figure}[h!]
	\begin{center}
		{\includegraphics[width=5.5in]{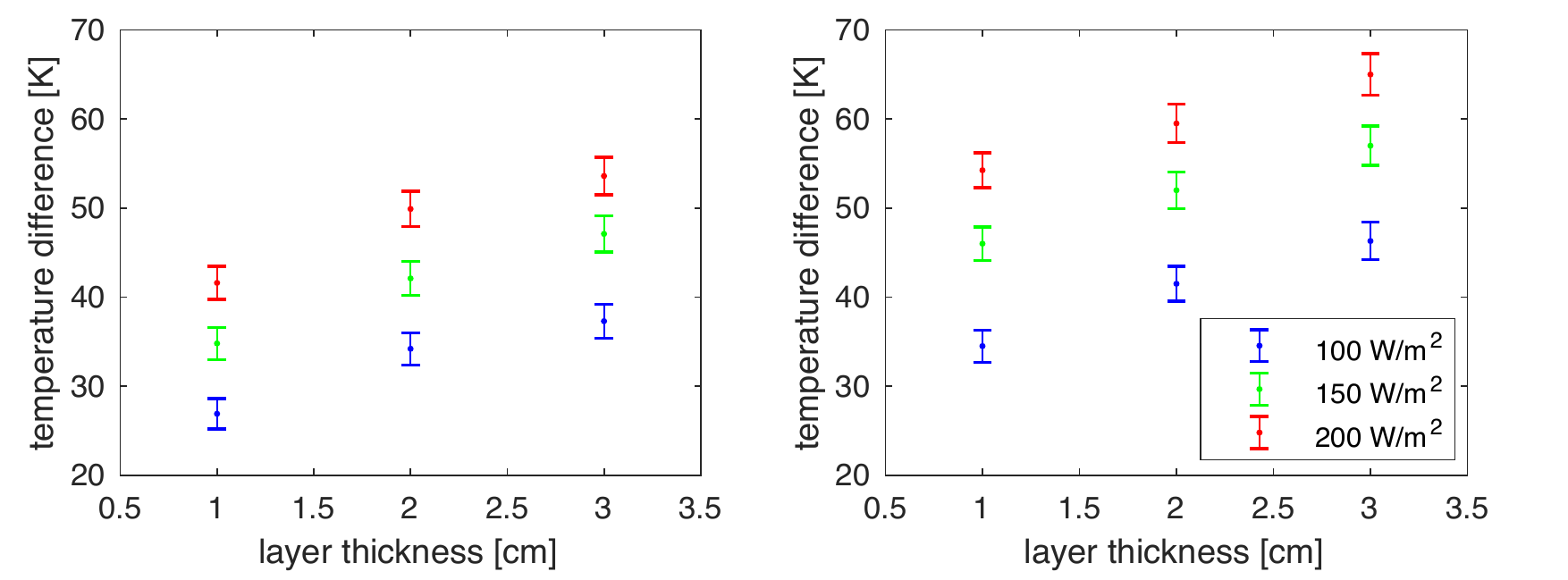}}
	\end{center}
	\caption{\textbf{Results of the silica aerogel solid-state greenhouse warming experiments.} Temperature differences between the surface and top of the layer are shown, for aerogel particles (left) and tiles (right), as a function of the layer thickness. Colors indicate data for different incident visible light fluxes. For reference, the annual mean flux on Mars between 45$^\circ$S and 45$^\circ$N varies from about 130 to 170~W/m$^2$, with diurnal mean values varying from 50 to 250~W/m$^2$ over the course of the martian year. Error bars indicate the estimated standard deviations of the measurements, which were calculated by combining uncertainties due to the thermistor calibration, data acquisition and signal digitization in quadrature.}
\label{fig:results}
\end{figure}

\begin{figure}[h!]
	\begin{center}
		{\includegraphics[width=4.5in]{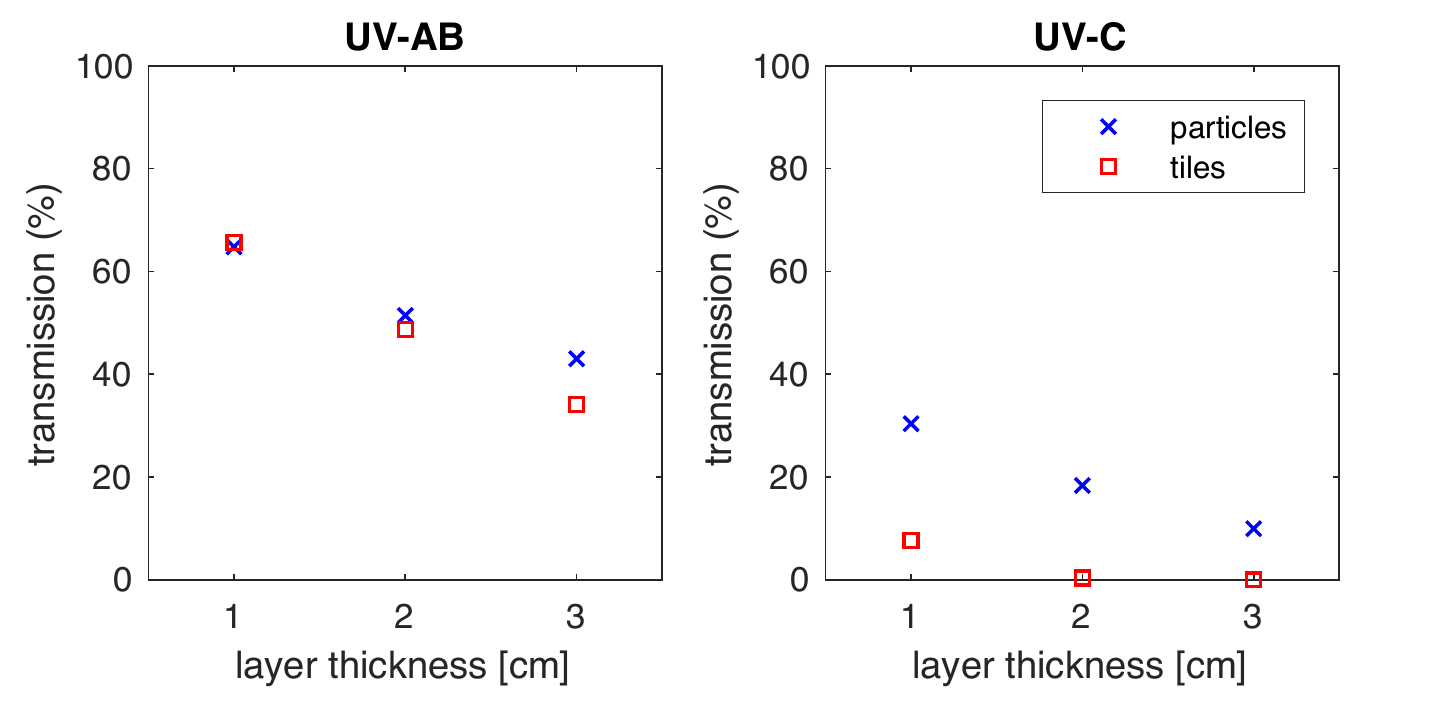}}
	\end{center}
	\caption{\textbf{Attenuation of ultraviolet radiation by silica aerogel.} Plot of UV-A/B (280-400~nm; left) and UV-C (220-275~nm; right) transmission by silica aerogel layers (particles and tiles) of thickness varying from 1- to 3-cm. Both particles and tiles attenuate UV-C effectively, with the transmission of UV-C through tile layers of 2-cm thickness or more dropping to below 0.5\%.}
\label{fig:transmission}
\end{figure}

\begin{figure}[h!]
	\begin{center}
		{\includegraphics[width=4.5in]{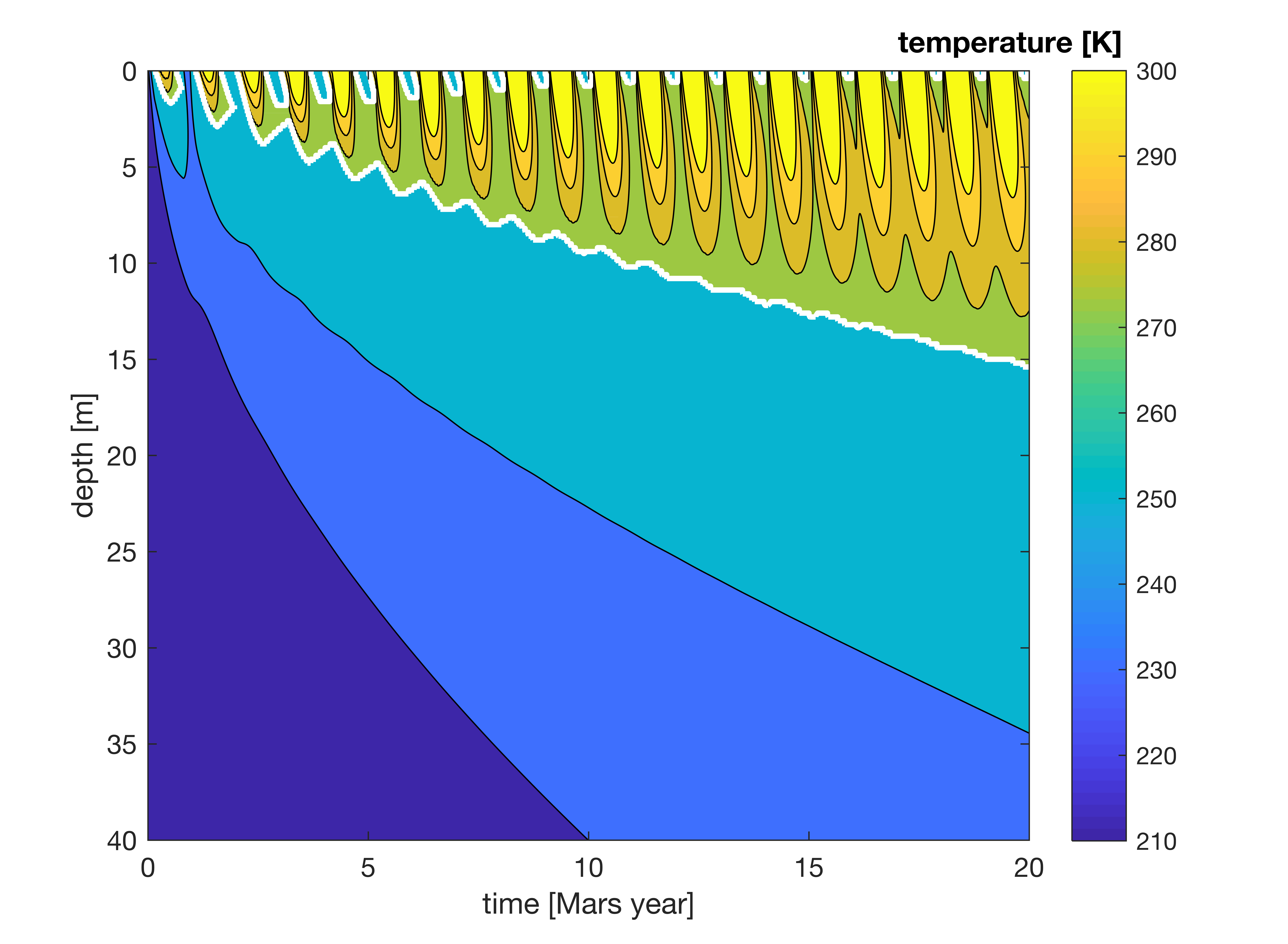}}
	\end{center}
	\caption{\textbf{Simulated warming underneath a silica-aerogel solid-state greenhouse habitat on Mars.} Contours show the sub-surface temperature evolution vs. time in an ice-rich regolith underneath a 2.5-cm silica aerogel layer on Mars in the Arabia Terra / Deuteronilus Mensae region (40$^\circ$N, 340$^\circ$W). The white contour shows the 273~K line corresponding to the melting point of pure water.}
\label{fig:sim_SSG}
\end{figure}

\clearpage

\section{Methods}

\subsection{Experimental}

Our experimental setup consisted of an {18~cm~$\times$~18~cm} solid-state greenhouse layer of variable thickness surrounded by polystyrene for insulation, with a solar simulator positioned above the layer to provide varying levels of visible irradiance. For the solid-state greenhouse layer, we used combinations of silica aerogel particles ({radii between 700~$\upmu$m and 4~mm}; Lumira\textsuperscript{\textregistered} from Cabot Aerogel) and tiles ({10~cm~$\times$~10~cm~$\times$~1~cm;} Large Hydrophobic Silica Tiles from Tiem Factory Inc.).  In the particle experiments, the entire layer was filled with particles, while in the tile experiments, the tiles  were placed in the centre and the remaining volume was filled with particles. Shallow pile black felt of visible albedo $<$ 0.01 (Protostar flocked light trap material) was placed below the aerogel layer to maximize the absorption of incoming visible radiation from the solar simulator.

Temperature data was collected via an array of compact (0.8~mm diameter) NTC thermistors. The thermistors were calibrated vs. a reference digital thermometer (Traceable\textsuperscript{\textregistered} Model 1235D30) between 0 and 100$^\circ$~C by suspending them in a continuously stirred water bath and recording the thermometer reading and thermistor resistance simultaneously. A least-squares fit was then used to determine resistance as a function of temperature according to the formula
\begin{equation}
R = C\left(\frac{T}{T_0}\right)^\beta\label{eq:therm}
\end{equation}
with $R$ resistance in k$\Omega$, $T$ temperature in K,  $T_0 = 273.15$~K a reference temperature, and $C$ and $\beta$ are calibration constants. We also tested the difference in resistance between individual thermistors at the same temperature and found it to be minimal compared to other error sources in the temperature range of interest.  Thermistors were attached to the base, top and exterior of the solid-state greenhouse apparatus and connected to a voltage divider circuit connected to a multiplexer / ESP8266 micro-controller for data acquisition. We also used a small thermal camera (Seek Thermal Imager) as an additional check on temperature data recorded by the thermistors and to diagnose side and base regions of elevated heat flux during the experimental setup.

Visible illumination to simulate the solar flux was provided by a 250~W protected pulse-start metal-halide lamp. A metal-halide light source was chosen because it approximates the zero-air-mass (AM0) solar spectrum more closely at most wavelengths than other light sources such as xenon arc \citemethods{Dong2015} at a lower cost and reduced explosion risk. The lamp was encased in a fan-cooled light box with mirrored ceiling and black sidewalls, to maximize the transmission of well-collimated light to the aerogel layer. A glass shield was placed between the lamp and experiment as a precaution against explosive lamp failure. The percentage of solar flux occuring above the wavelengths at which glass begins to absorb significantly (2-3~$\upmu$m) is of order a few percent, so absorption of near-infrared radiation by the shield was not judged to be a significant source of error in our results.  The optical properties of silica aerogel do not vary significantly across the visible wavelength range \citemethods{Fu2015}, so the relatively small differences between the metal-halide lamp spectrum and the solar spectrum incident at Mars' surface were not a significant source of uncertainty in our results either. The lamp was positioned between 20 and 30~cm above the silica aerogel sample. A lab jack was used to perform adjustments to the aerogel layer and pyranometer height in order to vary the received radiant flux. All experiments were run until thermal equilibrium was reached, which was judged by observing the value and rate of change of temperature at the base and top of the silica aerogel layer. Typically, this took around two hours for each experiment. Experiments were performed at ambient pressure and a background temperature of $298 \pm 1$~K.

The broadband visible fluxes incident on the sample were measured using a first class double glass dome pyranometer (Hukseflux Instruments model SR-11). The SR-11 model  used had a sensitivity of $14.22\times 10^{-6}\pm 0.15$~V/(W/m$^2$) and hence a $\pm 1\%$ calibration error. The estimated hourly uncertainty at equatorial latitudes during field observations stated in the pyranometer's technical literature were $\pm 3.1$~\%, which is the value we chose to use when stating our flux measurement uncertainties. Spatial variation of the incident flux was recorded by moving the pyranometer across an 0.15~m~$\times$~0.15~m grid in 3~cm intervals. Temporal variability of the flux was also measured and found to be $<2$~\% over a 2-hour interval. The ultraviolet transmission experiments were performed using a compact 4~W UV lamp with dual tubes to emit radiation peaking in either the 365~nm (UV-A/B) or 254~nm (UV-C) range. For UV measurement, we used calibration-certified Sper Scientific UV-A/B and UV-C detectors, which had a quoted accuracy of $\pm 4 \%$. The UV-A/B detector had peak sensitivity in the 350-360~nm range, with the calibration point at 365~nm, while the UV-C detector had peak sensitivity at 255-265~nm, with the calibration point at 254~nm. 

\subsection{Temperature Error Analysis}

For the temperature difference measurement, we considered errors from four sources: the data acquisition digitization error, the uncertainty in the voltage divider resistance, and the errors in the calibration parameters $C$ and $\beta$. The digitization error was 3.3~V/$2^{10}$ given the 10-bit data acquisition system used, or 3.2~mV, while the voltage divider resistance error was 1\%, or 0.1~k$\Omega$ for the 10~k$\Omega$ resistor. The errors in $C$ and $\beta$ were calculated from the log-linear least squares fit.

From equation~\eqref{eq:therm} and the voltage divider equation, the temperature difference between the base and top of the aerogel layer is
\begin{equation}
\Delta T(V_b,V_a) = T(V_b)- T(V_a) = T_0 \left[\left( \frac{ R_{1,b}V_b}{ C(V_0 - V_b)}\right)^{1/\beta} - \left( \frac{ R_{1,a}V_a}{ C(V_0 - V_a)}\right)^{1/\beta}\right]
\label{eq:Tdiff}
\end{equation}
where $V_0 = 3.3$~V is the peak voltage, $T(V_a)$ is the top-of-layer temperature, $T(V_b)$ is the base temperature, and $V_i$  and $R_{1,i}$ correspond to the output voltage and divider circuit fixed resistance for location $i$. We propagated uncertainties in $V_a$, $R_{1,a}$, $V_b$, $R_{1,b}$, $C$ and $\beta$ via a Taylor expansion assuming small uncertainties \citemethods{Barlow1989}. The resulting estimated uncertainties in $\Delta T$ were used to produce the error bars in Fig.~2 in the main text.

\subsection{Theory and Numerical Analysis}

\subsubsection*{Extreme upper limit of solid-state greenhouse warming potential}

An idealized upper limit to solid-state greenhouse warming can be derived by considering a material with zero thermal conductivity that is perfectly transparent below some cutoff wavelength $\lambda_c$ but absorbing at longer wavelengths. Under these circumstances, cooling can only occur via radiation from the base of the layer in the visible, and the base energy balance becomes 
\begin{equation}
\pi \int_0^{\lambda_c} B_\lambda[T] d\lambda = F_b, \label{eq:maxposs}
\end{equation}
where $\lambda$ is wavelength, $B_\lambda[T]$ is the Planck spectral irradiance, $T$ is the base temperature, and $F_b$ is the visible radiation (at wavelengths shorter than $\lambda_c$) absorbed at the layer base. Given the standard definition for $B_\lambda[T]$, \eqref{eq:maxposs} can be solved by a root-finding approach. The annual global mean solar flux received by Mars is around 150~W/m$^2$. Given $\lambda_c = 2$~$\upmu$m and $F_b = 150$~W/m$^2$, equation~\eqref{eq:maxposs} yields $T = 721$~K, which is close to the surface temperature of Venus. Shifting $\lambda_c$ to smaller values would lead to even higher $T$ values, with the temperature achieved asymptoting to the solar spectrum effective blackbody temperature as $\lambda_c \to 0$.

\subsubsection*{Optimal thickness of a solid-state greenhouse layer}

More realistically, we can determine the thickness required for a solid-state greenhouse layer to maximize surface temperature when its extinction optical depth in the visible is non-negligible, as is the case for all real materials. Here we neglect three-dimensional and basal conduction effects and assume constant thermal conductivity. We also assume that the solid-state greenhouse layer effectively absorbs infrared radiation, such that conduction is the dominant mode of heat transport in the layer. This analysis builds upon previous theoretical  studies of the solid-state greenhouse effect in snow and ice, including \citemain{Brown1987}, \citemethods{Matson1989} and \citemain{Pilorget2011}.

If the solid-state greenhouse layer has a non-zero extinction coefficient in the visible, the solar flux that reaches the base of the layer, $F_b$, will depend on $h$, the total layer thickness. Then, the total warming will depend on a balance between the attenuation of visible radiation and the thermal insulation provided by the layer. From the thermal diffusion equation, the steady-state temperature gradient inside the layer is \citemethods{Pierrehumbert2011BOOK}
\begin{equation}
\frac{dT}{dz} = -\frac{F_b(h)}{\kappa}. 
\end{equation}
where $\kappa$ is the solid-state greenhouse layer thermal conductivity and $z$ is the height in the layer. Integrating from 0 to $h$ yields
\begin{equation}
\Delta T = T_b - T_a = \frac{F_b(h) h}{\kappa}   \label{eq:deltaT_1}
\end{equation}
where $T_b$ and $T_a$ are the temperatures at the base and top of the layer, respectively. To find the peak temperature difference as a function of $h$, we differentiate to get
\begin{equation}
\frac{d \Delta T}{dh} = \frac{F_b(h)'h}{\kappa} + \frac{F_b(h)}{\kappa}.
\end{equation}
We then set $d \Delta T/dh$ to zero, yielding $F_b(h)'h = -F_b(h)$ (here the prime on $F_b(h)$ indicates differentiation with respect to $h$). Now if 
\begin{equation}
F_b(h) = F_a\E^{-\tau(h)/\overline{\mu}}= F_a\E^{-\alpha h/ \overline{\mu}}
\end{equation}
where $F_a$ is the incident solar flux on the layer, $\tau$ is the layer vertical path extinction optical depth, $\alpha$ is the layer extinction coefficient in the visible and $\overline \mu$ is the mean solar zenith angle cosine, it immediately follows that the optical depth for maximum warming $\tau_m$ is $\tau_m/\overline{\mu} = 1$ and the optimal layer height $h_m$ is 
\begin{equation}
h_{m} =\overline{\mu} / \alpha.
\end{equation}
Assuming vertical path transmission values of $\mathcal T = \E^{-\tau}=0.8$ and 0.6 for the 1~cm thick silica aerogel tile / particle layers, respectively, we find $\alpha=22.3$~m$^{-1}$ and  51.1~m$^{-1}$, or $h_m = 4.5$~cm and $2.0$~cm given $\overline \mu = 1$. The latter value is reasonably close to the layer depth of maximum warming achieved in the aerogel particle case (see Fig.~2 in the main text), with the slight difference likely due to multiple scattering effects.  From \eqref{eq:deltaT_1}, the theoretical maximum temperature difference is simply
\begin{equation}
\Delta T_{m} = \frac{ \overline{\mu} F_a\E^{-1}}{\alpha\kappa}.
\end{equation}
Given $F_a = 150$~W/m$^2$ and $\kappa = 0.02$~W/m/K, $\Delta T_{m} = 124$~K for the tiles and $54$~K for the particles. The temperature differences achieved in our experiments were somewhat lower than this because we only used aerogel layer thicknesses of up to 3~cm, and in addition losses from convection and sidewall and base conduction were non-negligible in our relatively small apparatus.

\subsubsection*{One-dimensional numerical model of solid-state greenhouse warming on Mars}

Our numerical model of the solid-state greenhouse effect on Mars calculates the diurnal average radiative transfer of the aerogel layer, the transport of heat via diffusion in the underlying regolith and the solar zenith angle as a function of time and location. Downwelling solar radiation and thermal radiation from the martian atmosphere are calculated using data from the Mars Climate Database (MCD) v5.3 Climatology scenario (\emph{http://www-mars.lmd.jussieu.fr/mars/access.html}) \citemain{Forget1999}. The solar zenith angle model is similar to that described in \citemethods{Macdonald2017} (see also \citemethods{Pierrehumbert2011BOOK}), with the martian orbital obliquity, eccentricity and season angle of perihelion taken into account via the method described in \citemethods{Hartmann2015}, and the season angle linked to time via Kepler's equation \citemethods{Prussing1993}. Our model output was validated under standard martian conditions vs. Fig.~1 of \citemethods{Levine1977} and by comparison with MCD results.  

Subsurface heat transport occurs in the model via thermal diffusion according to
\begin{equation}
c_h\rho\frac{\partial T}{\partial t} =  \frac{\partial}{\partial z}\left(\kappa_r \frac{\partial}{\partial z} T\right) + q(z) \label{eq:heat}
\end{equation}
Here $T$ is temperature, $t$ time, $z$ depth, $\kappa_r$ the regolith thermal conductivity, $c_h$ heat capacity, $\rho$ density and $q$ the local heating rate due to latent heat effects. We solve \eqref{eq:heat} via a centered difference in the spatial domain and explicit forward stepping in time.

Our model domain extends from the surface down to 80~m depth, and we integrate over a time period of 15 martian years. We neglect horizontal heat losses, so our model is appropriate for a setup where the horizontal extent of the solid-state greenhouse layer is tens of meters or more in both directions. As the solid-state greenhouse layer is only a few cm thick, we assume that it is in thermal equilibrium at every timestep. The thermal balance at the top of the layer is taken to be 
\begin{equation}
\sigma T_a^4 - F_{ir} = F_1 =  \kappa\frac{T_b - T_a}{h}
\end{equation}
with $\kappa$ the solid-state greenhouse thermal conductivity, $h$ the solid-state greenhouse layer thickness, $F_{ir}$ the downwelling thermal radiation from the martian atmosphere (supplied from the MCD) and $T_b$ the temperature immediately underneath the layer. This approach neglects additional heating or cooling of the surface due to sensible or latent atmospheric effects, which are of secondary importance due to the low density of Mars' atmosphere \citemain{Pilorget2011}.  This equation is solved for $T_a$ via a root-finding method at every timestep. $\kappa$ is taken to be $0.01$~W/m/K, which is an appropriate value for silica aerogel thermal conductivity given atmospheric pressures under 0.1~bar \citemain{Dorcheh2008}. The temperature just underneath the solid-state greenhouse layer is then evolved according to 
\begin{equation}
c_h \rho \frac{\partial  T_b}{\partial t} = (F_2 + F_{abs} - F_1)/\Delta z
\end{equation}
Here $F_2$ is the conductive heat flux from the layer below, $\Delta z$ is the subsurface numerical discretization thickness, and $F_{abs}$ is the absorbed solar flux. The absorbed solar flux is modeled as 
\begin{equation}
F_{abs} = F_{sol}\E^{-\tau/\overline \mu(t,\lambda)}.
\end{equation}
Here $F_{sol}$ is the diurnal mean solar flux at the surface of Mars (supplied from the MCD) and $\overline \mu(t,\lambda)$ is the diurnal mean solar zenith angle cosine output from the insolation model at time $t$ and latitude $\lambda$. Finally, $\tau = \alpha h$ is the solid-state greenhouse layer vertical path extinction optical depth, as in our previous analysis. {We conservatively neglect the diffuse visible flux to the layer base due to multiple scattering effects.}

The initial surface temperature in our simulation is taken to be the annual mean surface temperature in that location, based on MCD results. The initial regolith temperature gradient is set to the martian geotherm, which we take to be $15$~K/km based on an assumed geothermal heat flux of $F_{geo}=30$~mW/m$^2$  and mean regolith conductivity of 2~W/m/K, following \citemethods{Clifford1993}. At the bottom boundary, we assume a fixed heat flux equal to $F_{geo}$. 

We assume the regolith to be porous and saturated with ice, with the ice volume mixing ratio set to 0.5, corresponding to an ice-rich mid-latitude region such as Deuteronilus Mensae. Regolith density and sensible heat capacity were set by taking a weighted average of basalt and water ice. We found fairly low sensitivity of our results to the assumed ice/rock ratio. We take the latent heat during ice freezing and water melting into account in the thermal calculation by tracking the concentration of ice and water at each level through time, in a similar way to in \citemain{Pilorget2011}.  We then force the temperature to remain at or below 273.15~K whenever heat is entering a given layer and ice is still present, and assume that this heat is entirely used to melt the ice. Similar constraints are applied for the freezing of liquid water under cooling conditions.  Our numerical model has been validated vs. an analytic solution (propagation of a gaussian pulse) and verified to conserve energy and total \ce{H2O} mass globally to machine precision at every timestep. 

\textbf{Data availability statement:} The data that support the plots within this paper and other findings of this study are available from the corresponding author upon reasonable request.

\textbf{Code availability statement:} The one-dimensional solid-state greenhouse numerical model is available open-source at \emph{https://github.com/wordsworthgroup/Mars\_SSG\_2019}.

\end{document}